\documentclass[aps,prd,reprint,floatfix,longbibliography]{revtex4-2}

\usepackage[T1]{fontenc}
\usepackage[utf8]{inputenc}
\usepackage{amsmath,amssymb,bm}
\usepackage{booktabs}
\usepackage{graphicx}
\usepackage{array}
\usepackage{hyperref}
\usepackage{xcolor}

\hypersetup{
  colorlinks=true,
  linkcolor=blue,
  citecolor=blue,
  urlcolor=blue
}

\newcommand{\E}{\mathbf{E}}
\newcommand{\B}{\mathbf{B}}
\newcommand{\J}{\mathbf{J}}
\newcommand{\x}{\mathbf{x}}
\newcommand{\p}{\mathbf{p}}
\newcommand{\vvec}{\mathbf{v}}
\newcommand{\grad}{\nabla}
\newcommand{\dt}{\Delta t}

\newcommand{\ag}{g_{a\gamma\gamma}}
\newcommand{\ma}{m_a}
\newcommand{\Om}{\omega_0}

\begin{document}

\title{Coherent axion electrodynamics in OSIRIS: a counter-propagating axion--photon parametric instability}

\author{Miles Radford}
\email[Contact author: ]{miles.radford@physics.ox.ac.uk}
\affiliation{Department of Physics, University of Oxford, Parks Road, Oxford OX1 3PU, United Kingdom}

\author{Pablo Bilbao}
\affiliation{Department of Physics, University of Oxford, Parks Road, Oxford OX1 3PU, United Kingdom}

\author{Thomas Grismayer}
\affiliation{GoLP/Instituto de Plasmas e Fus\~ao Nuclear, Instituto Superior T\'ecnico, Universidade de Lisboa, Lisboa, Portugal}

\author{Luis O. Silva}
\affiliation{GoLP/Instituto de Plasmas e Fus\~ao Nuclear, Instituto Superior T\'ecnico, Universidade de Lisboa, Lisboa, Portugal}

\author{Subir Sarkar}
\affiliation{Department of Physics, University of Oxford, Parks Road, Oxford OX1 3PU, United Kingdom}

\author{Robert Bingham}
\affiliation{Department of Physics, University of Strathclyde, Glasgow G4 0NG, United Kingdom}
\affiliation{Rutherford Appleton Laboratory, STFC, Didcot OX11 0QX, United Kingdom}

\author{Gianluca Gregori}
\affiliation{Department of Physics, University of Oxford, Parks Road, Oxford OX1 3PU, United Kingdom}

\date{\today}

\begin{abstract}
We present a coherent axion-electrodynamics extension of the OSIRIS particle-in-cell code and use it to identify and benchmark a parametric axion--photon instability supported by two equal counter-propagating electromagnetic pumps. The module evolves a cell-centered axion field together with a separately stored first-order electromagnetic response on the native staggered OSIRIS grid. Re-examining the previously proposed strictly co-linear branch shows that it is removed when the axion--Maxwell equations are reduced consistently. A closely related standing-wave geometry, however, supports a distinct Floquet instability. We derive the coupled Floquet problem for an equal-amplitude counter-propagating pump pair and obtain a reduced analytic growth rate and cutoff for the dominant long-wavelength envelope branch. For $m_a/\omega_0=0.06$ and a per-beam vector-potential amplitude $A_L=0.3$, simulations at $g_{a\gamma\gamma}=0.60$, $0.80$, and $1.00$ reproduce both the growth-rate spectrum and the coupling-dependent instability boundary. The reduced theory predicts envelope-mode cutoffs $\ell_+=8.44$, $10.80$, and $13.23$, while OSIRIS finds exponential growth through $\ell=8$, $10$, and $13$, respectively, with the next resolved modes stable. The G060 electromagnetic diagnostics additionally resolve exponentially growing sidebands at twice the pump wavenumber while the zero-wavenumber electric component exhibits no sustained growth, as expected for the unstable Floquet eigenvector. The agreement provides an end-to-end benchmark of the coherent axion--electromagnetic coupling and demonstrates that parametric growth of the axion field is present in a two-beam pump geometry.
\end{abstract}

\maketitle

\section{Introduction}

Axions were originally proposed as the pseudo-Nambu--Goldstone bosons associated with the Peccei--Quinn solution to the strong-CP problem in quantum chromodynamics (QCD) \cite{Peccei1977PRL,Peccei1977PRD,Weinberg1978,Wilczek1978}. More general axion-like particles (ALPs) retain similar low-energy interactions without satisfying the strict QCD axion mass--coupling relation, motivating laboratory, astrophysical, haloscope, helioscope, polarimetry, and light-shining-through-wall searches over a wide parameter space \cite{Kim2010,DiLuzio2020,Graham2015,Irastorza2018,Sikivie2021,CAST2017,ADMX2018,ALPS2010,Halliday2025,Heaton2026}.

For laser and plasma applications, the relevant interaction is the pseudoscalar axion--photon coupling
\begin{equation}
  \mathcal{L}_{a\gamma\gamma}=-\ag a\,\E\cdot\B,
  \label{eq:lint}
\end{equation}
where $a$ is the axion field and $\ag$ is the axion--photon coupling. We use the $a$ convention throughout; the scalar variable $\phi$ used in Ref.~\cite{An2024} has the opposite sign, $\phi=-a$. In natural units, the coupled equations, including ordinary charge and current sources, are \cite{Sikivie1983,Wilczek1987AxionED,Sikivie2021}
\begin{align}
  \partial_t^2 a-\grad^2 a+\ma^2 a &= -\ag\E\cdot\B, \label{eq:kg-cont}\\
  \grad\cdot\E &= \rho+\ag\B\cdot\grad a, \label{eq:gauss-cont}\\
  \grad\cdot\B &=0, \label{eq:divb-cont}\\
  \grad\times\E &= -\partial_t\B, \label{eq:faraday-cont}\\
  \grad\times\B &= \partial_t\E+\J
  +\ag\left[\E\times\grad a-(\partial_t a)\B\right].
  \label{eq:ampere-cont}
\end{align}
Here $\rho$ is the ordinary charge density and $\J$ is the ordinary current density. These equations are particularly suitable for particle-in-cell (PIC) simulation because the ordinary electromagnetic and plasma degrees of freedom are already evolved self-consistently, while the coherent scalar field can be added as an additional grid quantity.

Ref.~\cite{An2024} demonstrated a field-perturbation-splitting (FPS) implementation of axion electrodynamics in the EPOCH PIC code \cite{Arber2015}. We follow the same numerical principle in OSIRIS \cite{Fonseca2002,OSIRISGitHub}: the large zeroth-order electromagnetic field is retained in the native OSIRIS arrays, while the much smaller axion-induced first-order field is evolved separately. This prevents a weak coherent response from being lost numerically beneath a large laser field. The present coherent-field module complements our earlier OSIRIS axion macroparticle and Monte-Carlo framework for incoherent emission, absorption, transport, and detailed balance in thermal plasmas \cite{Radford2026MCC}.

The original motivation for the present benchmark was the co-linear parametric instability proposed in Ref.~\cite{Beyer2023}. During validation we re-derived that reduction directly from Eqs.~(\ref{eq:kg-cont})--(\ref{eq:ampere-cont}) and found that an algebraic step in the co-linear wave equation is inconsistent with the full axion--Maxwell system; retaining the corresponding term removes that particular exponential branch. We therefore do not use the strictly co-linear configuration as a benchmark here. Instead, we consider the nearby geometry of two equal, counter-propagating pump waves. Their standing-wave structure supplies additional spatial harmonics and the corrected equations then admit a distinct parametric instability.

The resulting interaction is naturally described as a Floquet sideband problem rather than as a single Stokes three-wave decay. Both pump directions and multiple electromagnetic sidebands participate, in close analogy with the sideband structure familiar from parametric laser--plasma instabilities \cite{Forslund1973,Drake1974,Kruer1988}. We derive the full coupled Floquet system, obtain a compact analytic limit for the dominant branch, and compare both its growth rate and its mode cutoff directly with OSIRIS simulations. The agreement is sufficiently accurate that the same instability can serve as a stringent end-to-end test of the coherent axion-electrodynamics implementation.

\section{Counter-propagating instability theory}
\label{sec:two-beam-theory}

\subsection{Linearized equations}

We work in one spatial dimension $z$ and choose two equal counter-propagating pumps with a common linear polarization. Their combined vector potential is written
\begin{equation}
  \mathbf{A}_0=P(z,t)\,\hat{\mathbf{x}},
  \qquad
  P=-2A_L\sin(\Om t)\cos(k_0 z),
  \label{eq:pump-potential}
\end{equation}
with $k_0=\Om$ in vacuum and $A_L$ the vector-potential amplitude of each travelling wave. The corresponding pump fields are
\begin{align}
  \mathbf{E}_0 &=-\partial_tP\,\hat{\mathbf{x}},\\
  \mathbf{B}_0 &= \partial_zP\,\hat{\mathbf{y}}.
\end{align}
We take the axion-induced electromagnetic daughter to have orthogonal polarization,
\begin{equation}
  \mathbf{A}_1=b(z,t)\,\hat{\mathbf{y}},
  \qquad
  \mathbf{E}_1=-\partial_t b\,\hat{\mathbf{y}},
  \qquad
  \mathbf{B}_1=-\partial_z b\,\hat{\mathbf{x}}.
  \label{eq:daughter-potential}
\end{equation}
Linearizing Eqs.~(\ref{eq:kg-cont})--(\ref{eq:ampere-cont}) in $a$ and $b$ about the prescribed pump gives
\begin{align}
  (\partial_t^2-\partial_z^2)b
  &=\ag\left(P_t a_z-P_z a_t\right),
  \label{eq:coupled-b}\\
  (\partial_t^2-\partial_z^2+\ma^2)a
  &=\ag\left(P_z b_t-P_t b_z\right).
  \label{eq:coupled-a}
\end{align}
These two equations are the starting point for the instability analysis. In particular, Eq.~(\ref{eq:coupled-b}) retains both the spatial-gradient and time-derivative contributions required by the axion-modified Amp\`ere equation. For a single exactly co-linear travelling pump their combination does not generate the previously reported unstable low-frequency branch; the standing-wave pump in Eq.~(\ref{eq:pump-potential}) changes the coupling structure by adding the opposite-$k_0$ travelling-wave harmonic.

\begin{figure*}[!t]
  \centering
  \includegraphics[width=0.98\textwidth]{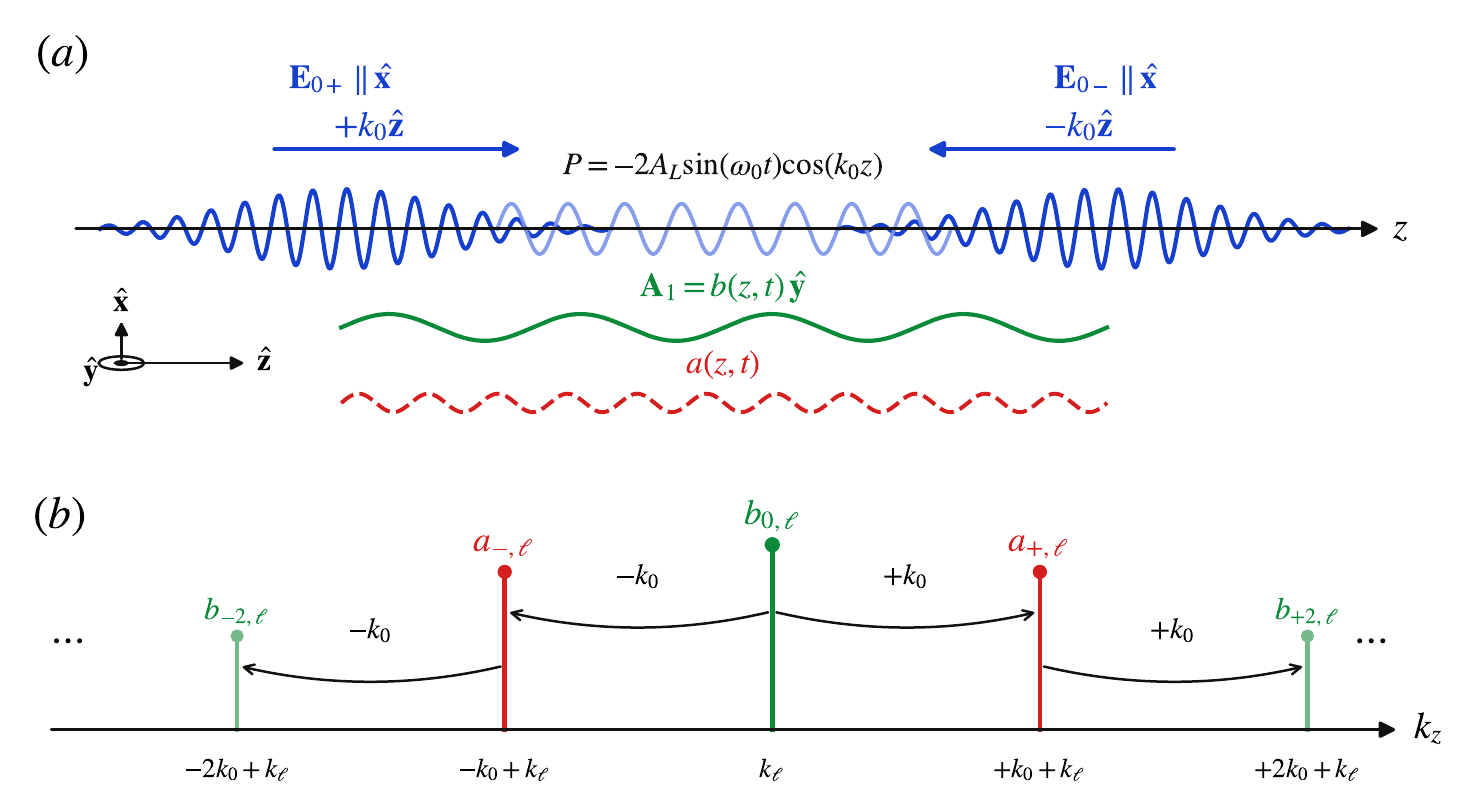}
  \caption{Schematic of the counter-propagating instability and its Floquet coupling structure, using the same coordinate convention as the analytic derivation. (a) Two equal pumps with wavevectors $+k_0\hat{\mathbf z}$ and $-k_0\hat{\mathbf z}$ and common vector-potential polarization $\hat{\mathbf x}$ form the standing-wave background $P=-2A_L\sin(\omega_0 t)\cos(k_0z)$. The axion-induced electromagnetic response has vector potential $\mathbf A_1=b\hat{\mathbf y}$, while $a$ is the scalar axion perturbation. The drawing is schematic and does not represent a single pump decaying into a fixed pair of daughters. (b) Signed-wavenumber view of one Floquet ladder for small envelope wavenumber $k_\ell$. The central electromagnetic harmonic $b_{0,\ell}$ at $k_\ell$ is coupled by the two pump harmonics to axion harmonics $a_{-,\ell}$ and $a_{+,\ell}$ at $-k_0+k_\ell$ and $+k_0+k_\ell$, respectively. Because the real scalar field contains conjugate $\pm k$ Fourier components, these two signed harmonics correspond to the positive absolute axion mode pair $N_0-\ell$ and $N_0+\ell$ used in the simulation analysis. Repeated action of the pump harmonics generates the neighboring electromagnetic components at $k_\ell\pm2k_0$ and the continuing alternating Floquet ladder indicated by the ellipses. The diagram shows allowed couplings, not relative eigenvector amplitudes: at exactly $k_\ell=0$, the unstable branch has $b_{00}=0$ even though its higher electromagnetic sidebands are nonzero. The schematic was prepared with Python graphics code developed with AI assistance.}
  \label{fig:counterprop-geometry}
\end{figure*}

It is useful to resolve the pump into its two travelling components,
\begin{equation}
  P=-A_L\sum_{s=\pm1}\sin(\Om t-s\Om z),
\end{equation}
and define
\begin{equation}
  \mathcal D_s=\Om\partial_z+s\Om\partial_t.
\end{equation}
Equations~(\ref{eq:coupled-b}) and (\ref{eq:coupled-a}) then become
\begin{align}
  (\partial_t^2-\partial_z^2)b
  &=-\ag A_L\sum_{s=\pm1}
  \cos(\Om t-s\Om z)\,\mathcal D_s a,
  \label{eq:b-operator}\\
  (\partial_t^2-\partial_z^2+\ma^2)a
  &=+\ag A_L\sum_{s=\pm1}
  \cos(\Om t-s\Om z)\,\mathcal D_s b.
  \label{eq:a-operator}
\end{align}
The explicit periodicity in both space and time makes this a Floquet--Bloch problem. The geometry and the resulting sideband coupling are summarized in Fig.~\ref{fig:counterprop-geometry}.

\subsection{Floquet system}

We expand both perturbations in harmonics of the pump. Here $n,m\in\mathbb{Z}$ are the temporal and spatial Floquet indices, while $\omega$ and $k$ are the corresponding quasi-frequency and quasi-wavenumber within the reduced Floquet zone:
\begin{align}
 a(z,t)&=\sum_{n,m}a_{nm}
 e^{i(k+m\Om)z-i(\omega+n\Om)t},\\
 b(z,t)&=\sum_{n,m}b_{nm}
 e^{i(k+m\Om)z-i(\omega+n\Om)t}.
 \label{eq:floquet-expansion}
\end{align}
For a harmonic $(n,m)$ define
\begin{align}
 D_{\gamma}^{nm}&=(\omega+n\Om)^2-(k+m\Om)^2,\\
 D_a^{nm}&=(\omega+n\Om)^2-(k+m\Om)^2-\ma^2,\\
 Q_s^{nm}&=\Om\left[(k+m\Om)-s(\omega+n\Om)\right].
\end{align}
Multiplication by a pump harmonic shifts the Floquet indices as
$(n,m)\rightarrow(n+\eta,m+\eta s)$, where $s=\pm1$ labels the two pump propagation directions and $\eta=\pm1$ labels the positive- and negative-frequency complex harmonics of each cosine. Substitution of Eq.~(\ref{eq:floquet-expansion}) into Eqs.~(\ref{eq:b-operator}) and (\ref{eq:a-operator}) therefore gives the sparse coupled system
\begin{align}
 D_{\gamma}^{nm}b_{nm}
 &=\frac{i\ag A_L}{2}
 \sum_{s=\pm1}\sum_{\eta=\pm1}
 Q_s^{n+\eta,m+\eta s}
 a_{n+\eta,m+\eta s},
 \label{eq:floquet-b}\\
 D_a^{nm}a_{nm}
 &=-\frac{i\ag A_L}{2}
 \sum_{s=\pm1}\sum_{\eta=\pm1}
 Q_s^{n+\eta,m+\eta s}
 b_{n+\eta,m+\eta s}.
 \label{eq:floquet-a}
\end{align}
The complex roots $\omega(k)$ of the truncated Floquet determinant determine the linear modes; $\operatorname{Im}\omega>0$ corresponds to exponential growth. For the theory curves used below we solve Eqs.~(\ref{eq:floquet-b})--(\ref{eq:floquet-a}) numerically with a symmetric truncation $|n|,|m|\le6$. Increasing the truncation leaves the dominant branch unchanged at the plotting resolution.

The instability is centered on axion sidebands whose absolute Fourier mode numbers lie close to the pump mode. For a periodic domain of longitudinal length $L$, we therefore write the physical axion Fourier mode numbers as $N_a=N_0\pm\ell$, where $N_0=Lk_0/(2\pi)$ is the integer pump mode number and $\ell$ is the non-negative envelope-mode index. The corresponding envelope wavenumber is
\begin{equation}
  k_{\ell}=\frac{2\pi\ell}{L}.
  \label{eq:envelope-k}
\end{equation}
This is the $k$ that enters the Floquet eigenproblem above.

\subsection{Reduced analytic growth rate and cutoff}

For $\mu=\ma/\Om\ll1$, the dominant long-envelope sector can be reduced in a controlled asymptotic limit. We assume $\alpha=\ag A_L=\mathcal O(\mu)$ and $\omega/\Om,k/\Om=\mathcal O(\mu^2)$. The retained central photon harmonic is $B=b_{00}$, coupled to the four resonant axion harmonics $a_1=a_{-1,-1}$, $a_2=a_{-1,+1}$, $a_3=a_{+1,-1}$ and $a_4=a_{+1,+1}$. To the same leading order, the off-resonant photon harmonics $b_{-2,0}$, $b_{0,-2}$, $b_{0,+2}$ and $b_{+2,0}$ can be eliminated; their contributions to the axion block must be retained. Define
\begin{align}
 \Omega&=\frac{\omega\Om}{\ma^2}, &
 K&=\frac{k\Om}{\ma^2}, \notag\\
 r&=\left(\frac{\ag A_L\Om}{\ma}\right)^2, &
 \lambda&=\frac{\ag A_L\Om}{2\ma},
 \label{eq:reduced-scaling}
\end{align}
where $\Omega$ and $K$ are the scaled envelope frequency and wavenumber, $r=4\lambda^2$ is the dimensionless pump-strength parameter, and $\lambda$ retains the sign of the coupling. Setting $x_j=a_j/\mu$ for $j=1,\ldots,4$, and $S=\Omega^2-K^2$, the resulting homogeneous equations are $\mathsf M\,(x_1,x_2,x_3,x_4,B)^T=0$ with
\begin{widetext}
\begin{equation}
\mathsf M=\begin{pmatrix}
 2K-2\Omega-1 & r/4 & r/4 & 0 & i\lambda(K-\Omega)\\
 r/4 & -2K-2\Omega-1 & 0 & r/4 & i\lambda(K+\Omega)\\
 r/4 & 0 & 2K+2\Omega-1 & r/4 & i\lambda(K+\Omega)\\
 0 & r/4 & r/4 & -2K+2\Omega-1 & i\lambda(K-\Omega)\\
 -i\lambda(K-\Omega) & -i\lambda(K+\Omega) & -i\lambda(K+\Omega) & -i\lambda(K-\Omega) & S
\end{pmatrix}.
\label{eq:five-mode-matrix}
\end{equation}
\end{widetext}
Let $H$ denote the upper-left $4\times4$ axion block of $\mathsf M$ and $\mathbf u=(K-\Omega,K+\Omega,K+\Omega,K-\Omega)^T$. A direct block-determinant expansion gives the \emph{polynomial identity}
\begin{equation}
 \det\mathsf M=S\det H-\frac r4\,\mathbf u^T\operatorname{adj}(H)\mathbf u,
 \label{eq:full-block-determinant}
\end{equation}
where $\operatorname{adj}(H)$ is the adjugate of $H$. No inverse $H^{-1}$ is introduced; Eq.~(\ref{eq:full-block-determinant}) remains valid even at $\det H=0$. Expanding $\det\mathsf M=0$ yields
\begin{align}
 0={}&64S^3-16(2+r)S^2 \notag\\
 &+\left[4(1+r)-3r^2-64K^2\right]S+8rK^2.
 \label{eq:scaled-cubic}
\end{align}
Equivalently, with $\varpi=\omega/\Om$, $\kappa=k/\Om$ and $Z=\varpi^2-\kappa^2$, the unscaled dispersion polynomial is
\begin{align}
 0={}&64Z^3-(32\mu^4+16\mu^2\alpha^2)Z^2 \notag\\
 &+\left(4\mu^8+4\mu^6\alpha^2-64\mu^4\kappa^2
 -3\mu^4\alpha^4\right)Z
 +8\mu^6\alpha^2\kappa^2.
 \label{eq:reduced-cubic}
\end{align}

At $K=0$, Eq.~(\ref{eq:scaled-cubic}) factorises, giving $S=0$, $S_+=(3r+2)/8$ and $S_-=(2-r)/8$. The growing solution for $r>2$ belongs to the latter root, at which $\det H=0$. It is nevertheless a genuine mode of the \emph{full} system: the corresponding eigenvector obeys
\begin{equation}
 B=0,\qquad (x_1,x_2,x_3,x_4)=(X,X,Y,Y),
 \label{eq:unstable-eigenvector}
\end{equation}
with nonzero complex amplitudes $X$ and $Y$ satisfying
\begin{equation}
 \begin{pmatrix}
 -1-2\Omega+r/4 & r/4\\
 r/4 & -1+2\Omega+r/4
 \end{pmatrix}\binom{X}{Y}=0.
 \label{eq:axion-reduced-matrix}
\end{equation}
Its determinant again gives $\Omega^2=(2-r)/8$. Thus the central low-frequency photon is absent for the exact $K=0$ growing eigenmode, whereas the off-resonant electromagnetic sidebands retained in the reduction need not vanish. The condition $r>2$ refers specifically to this zero-envelope-wavenumber branch and does not assert global stability of every Floquet sector for $r\le2$.

For the purely growing branch we write $\Gamma=\operatorname{Im}\omega>0$. Expanding its eigenfrequency around $K=0$ gives
\begin{align}
  \Gamma^2(k)
  &=\Gamma_0^2-C(r)k^2
  +\mathcal{O}\!\left(\frac{k^4\Om^2}{\ma^4}\right),
  \label{eq:standing-growth}\\
  \Gamma_0^2
  &=\frac{(\ag A_L)^2\ma^2}{8}
  -\frac{\ma^4}{4\Om^2},
  \label{eq:gamma0}\\
  C(r)&=\frac{r^2-6r+4}{r(r-2)}.
  \label{eq:small-k-coeff}
\end{align}
In the asymptotic strong-pump limit $r\gg1$, $C(r)\rightarrow1$, recovering the simpler approximation $\Gamma^2\simeq\Gamma_0^2-k^2$. The instability boundary of the reduced determinant is obtained separately by setting $\omega=0$, for which $S=-K^2$ and Eq.~(\ref{eq:scaled-cubic}) factorises as
\begin{equation}
  K^2\left(-8K^2+r+2\right)
  \left(8K^2+3r-2\right)=0.
  \label{eq:cutoff-factorization}
\end{equation}
Following the negative-$\Omega^2$ root continuously from $K=0$ shows that it reaches $\Omega=0$ at the positive root $K_+^2=(r+2)/8$; the other nonzero factor gives no positive $K^2$ in this regime. Thus the upper cutoff of this branch is
\begin{equation}
  k_+=\frac{\ma^2}{\Om}
  \sqrt{\frac{2+r}{8}}.
  \label{eq:standing-cutoff}
\end{equation}
For a periodic domain, Eq.~(\ref{eq:envelope-k}) gives
\begin{equation}
  \ell_+=\frac{L\ma^2}{2\pi\Om}
  \sqrt{\frac{2+r}{8}}.
  \label{eq:ell-cutoff}
\end{equation}
The full Floquet calculation is used for the quantitative theory curves in Sec.~\ref{sec:benchmark-results}; Eqs.~(\ref{eq:standing-growth})--(\ref{eq:small-k-coeff}) describe the local long-wavelength behavior, while Eqs.~(\ref{eq:standing-cutoff}) and (\ref{eq:ell-cutoff}) give the instability boundary of the reduced determinant.

\section{Implementation methodology}

\subsection{Overview of the OSIRIS integration}

The implementation introduces a new OSIRIS simulation subtype for coherent axion-electrodynamics calculations. This subtype extends the existing QED-capable OSIRIS simulation class rather than replacing it. The inherited OSIRIS machinery still performs the ordinary electromagnetic field advance, particle push, current deposition, boundary exchange, smoothing, laser injection, moving-window update, and diagnostics. The axion-electrodynamics extension adds a coherent axion-field state and an auxiliary first-order current state to this inherited loop.

The axion-field state stores the scalar axion field, its adjacent time level, diagnostic arrays, and perturbative electromagnetic fields. The auxiliary first-order current has the same general role for the perturbative electromagnetic update as the native OSIRIS current has for the ordinary Maxwell update. In this sense the implementation is a hybrid extension of the normal PIC algorithm: OSIRIS advances the zeroth-order plasma and electromagnetic fields through the standard PIC loop, while the axion module advances a coherent scalar field and the first-order electromagnetic response sourced by it.

Input is provided through an axion-field block in the OSIRIS deck. The main physical and numerical controls are the axion mass, the code-level axion-photon coupling, the scalar-field initialization mode, the Fourier-mode or pulse parameters used to initialize the axion field, the six directional boundary-condition strings, and the diagnostic report list. The axion-field input parser defaults the axion boundaries to periodic in active directions before applying user overrides. This default is useful for the simplest instability benchmark because a periodic domain provides a clean mapping between an input mode number and a resolved Fourier wavenumber.

\subsection{Field perturbation separation}

The continuum equations in Eqs.~(\ref{eq:kg-cont})--(\ref{eq:ampere-cont}) contain terms proportional to the product of the axion coupling and the electromagnetic fields. For realistic couplings, the electromagnetic response induced by the axion field can be extremely small compared with the background laser or plasma fields. The formulation of Ref.~\cite{An2024} addresses this by writing the electromagnetic fields, charge density, and current density as expansions in the axion-induced perturbation,
\begin{align}
  \E &= \E_0+\E_1, &
  \B &= \B_0+\B_1, \nonumber\\
  \rho &= \rho_0+\rho_1, &
  \J &= \J_0+\J_1,
  \label{eq:fps-split}
\end{align}
where the zeroth-order fields satisfy the ordinary Maxwell--PIC system and the first-order fields carry the axion-induced electromagnetic response \cite{An2024}. Retaining the first-order axion-induced fields separately gives
\begin{align}
  \grad\cdot\E_1 &= \rho_1 + \ag\B_0\cdot\grad a, \label{eq:fps-gauss}\\
  \grad\cdot\B_1 &= 0, \label{eq:fps-divb}\\
  \grad\times\E_1 &= -\partial_t\B_1, \label{eq:fps-faraday}\\
  \grad\times\B_1 &= \partial_t\E_1 + \J_1
  +\ag\left[\E_0\times\grad a-(\partial_t a)\B_0\right].
  \label{eq:fps-ampere}
\end{align}
These are the equations that motivate the numerical splitting used in OSIRIS.

In the OSIRIS implementation, the native electromagnetic field arrays store the zeroth-order fields $\E_0$ and $\B_0$. The axion module stores the perturbative electromagnetic fields $\E_1$ and $\B_1$ separately. This separation is not only a matter of convenience. It ensures that a small axion-induced field is represented in its own floating-point storage rather than as a tiny correction to a much larger laser field. The same logic is applied to the current response: the ordinary deposited current remains in the usual OSIRIS current state, while the first-order current is stored and advanced separately.

\subsection{Scalar axion-field storage and staggering}

The scalar axion field is stored on two adjacent half-time levels. We denote the discrete integer time index by $q$, such that $t^q=q\Delta t$, with the two stored scalar fields evaluated at $t^{q-1/2}$ and $t^{q+1/2}$. These should be interpreted operationally as neighboring time levels of the same scalar field, rather than as a continuum canonical field and canonical momentum pair. Quantities required at the intervening integer time $t^q$ are reconstructed from these two stored fields. The centered scalar field at this time is
\begin{equation}
  a^q =
\frac{1}{2}
\left(
a^{q-1/2}+a^{q+1/2}
\right),
  \label{eq:a-centered}
\end{equation}
while its centered time derivative is
\begin{equation}
(\partial_t a)^q =
\frac{a^{q+1/2}-a^{q-1/2}}{\Delta t}.
  \label{eq:adot-centered}
\end{equation}
These reconstructed quantities are used in the axion-induced
electromagnetic source terms and in the corresponding diagnostics.

Spatially, the scalar field is cell-centered. The native electromagnetic fields and the perturbative electromagnetic fields retain the usual Yee staggering. Consequently, all products such as $\E\cdot\B$, $(\partial_t a)\B_0$, and $\E_0\times\grad a$ require local interpolation or averaging to place the participating quantities on a common grid location. This is the same basic issue described in Ref.~\cite{An2024}, where the scalar field is located at the cell center and the electromagnetic components are staggered on the cell faces and edges.

\subsection{Finite-difference scalar-field update}

The free scalar-field update is a second-order centered finite-difference update of the Klein-Gordon equation. In three dimensions, the discrete Laplacian applied to the half-time scalar field is
\begin{align}
  \grad_h^2 a_{i,j,k}^{q+1/2}
  =&\frac{a_{i+1,j,k}^{q+1/2}-2a_{i,j,k}^{q+1/2}+a_{i-1,j,k}^{q+1/2}}{\Delta x^2}
  \nonumber\\
  &+\frac{a_{i,j+1,k}^{q+1/2}-2a_{i,j,k}^{q+1/2}+a_{i,j-1,k}^{q+1/2}}{\Delta y^2}
  \nonumber\\
  &+\frac{a_{i,j,k+1}^{q+1/2}-2a_{i,j,k}^{q+1/2}+a_{i,j,k-1}^{q+1/2}}{\Delta z^2} .
  \label{eq:discrete-laplace}
\end{align}
Suppressing the cell index on the updated field, the free update is then
\begin{align}
  a^{q+3/2}
  =&\,2a^{q+1/2}-a^{q-1/2}
  \nonumber\\
  &+\dt^2\left(\grad_h^2 a^{q+1/2}
  -\ma^2 a^{q+1/2}\right).
  \label{eq:free-kg-update}
\end{align}
The driven update adds the $\E\cdot\B$ source evaluated at the scalar-field location. In the $a$ convention adopted above, the source term is implemented as
\begin{align}
  a^{q+3/2}
  =&\,2a^{q+1/2}-a^{q-1/2}
  \nonumber\\
  &+\dt^2\left[\grad_h^2 a^{q+1/2}
  -\ma^2 a^{q+1/2}
  -\ag(\E\cdot\B)_c\right].
  \label{eq:driven-kg-update}
\end{align}
Here $(\E\cdot\B)_c$ is formed at the scalar cell center from the native OSIRIS fields plus the perturbative fields.

For the cell-centered source diagnostic, the electric components are averaged to the cell center. For example, in three dimensions the code uses the local Yee-grid averages
\begin{align}
  E_{x,c} &= \frac{1}{2}\left[E_x(i,j,k)+E_x(i-1,j,k)\right],\label{eq:ex-center}\\
  E_{y,c} &= \frac{1}{2}\left[E_y(i,j,k)+E_y(i,j-1,k)\right],\label{eq:ey-center}\\
  E_{z,c} &= \frac{1}{2}\left[E_z(i,j,k)+E_z(i,j,k-1)\right],\label{eq:ez-center}
\end{align}
with analogous four-point averages for magnetic components. The same centering is applied to the perturbative electromagnetic fields before adding them to the native fields in the $\E\cdot\B$ diagnostic. This centering is essential because the scalar source is a product of quantities stored at different Yee-grid locations.

Initial conditions for the scalar field are controlled from the input
deck. For periodic benchmarks, the axion field can be seeded with sinusoidal longitudinal Fourier modes specified by their integer mode number, amplitude, and phase. In a periodic domain of length $L_x$, an
integer mode number $n$ corresponds to the longitudinal wavenumber
\begin{equation}
  k_n = \frac{2\pi n}{L_x},
  \label{eq:box-mode}
\end{equation}
providing a direct correspondence between an OSIRIS Fourier mode and the wavenumber used in the Floquet analysis below.

The scalar solver uses a centered second-order leapfrog scheme and stores the axion field at two adjacent half-time levels, $a^{q-1/2}$ and $a^{q+1/2}$. Consequently, when an initial axion mode is specified at
the nominal time $t=0$, the values stored at the two neighboring
half-time levels must be chosen consistently with the time evolution of
that mode. Simply assigning the same specified $t=0$ amplitude without
accounting for the temporal staggering would introduce a small
initialization error and hence an artificial transient.

The required temporal evolution is determined using the finite-difference dispersion relation of the scalar solver. In the implementation discussion we denote the longitudinal simulation coordinate by $x$ (the first Cartesian coordinate used by OSIRIS); this is the same physical direction denoted by $z$ in the analytic derivation of Sec.~\ref{sec:two-beam-theory}. We use $d$ to label the spatial coordinate directions, with $d\in\{x,y,z\}$, $\Delta x_d$ denoting the grid spacing in direction $d$, and $k_d$ the corresponding Cartesian component of the wavevector. Substitution of a discrete plane wave into the Klein--Gordon update gives
\begin{equation}
\begin{split}
\left[\frac{2}{\Delta t}\sin\left(\frac{\omega_h\Delta t}{2}\right)\right]^2
={}&m_a^2\\
&+\sum_d
\left[\frac{2}{\Delta x_d}\sin\left(\frac{k_d\Delta x_d}{2}\right)\right]^2,
\end{split}
\end{equation}
where $\omega_h$ is the numerical frequency of the corresponding Fourier mode on the discrete grid. The spatial-direction index $d$ is distinct from the longitudinal Fourier-mode index $n$ introduced above; for the one-dimensional longitudinal seeds used in the present benchmark, $k_x=k_n=2\pi n/L_x$ and the transverse wavevector components vanish. In the continuum limit, $\Delta t,\Delta x_d\rightarrow0$, this reduces to the usual Klein--Gordon dispersion relation, $\omega^2=m_a^2+|\mathbf{k}|^2$.

For the standing-wave seeds used here, the initial condition is chosen
to have zero centered time derivative at $t=0$. A mode with centered
amplitude $a_0\sin(k_n x+\theta)$, where $a_0$ is the specified mode amplitude and $\theta$ its initial phase, is therefore represented on the two
stored half-time levels as
\begin{equation}
  a^{+1/2}(x)=a^{-1/2}(x)
  =
  a_0
  \cos\left(\frac{\omega_h\dt}{2}\right)
  \sin(k_n x+\theta).
\end{equation}
The factor $\cos(\omega_h\dt/2)$ is therefore not a modification of the
physical seed amplitude or dispersion. It accounts only for the fact
that the numerical solver stores the field at times displaced by
$\pm\dt/2$ from the nominal initial time. When several Fourier modes are
seeded, the correction is evaluated separately for each $k_n$ using its
corresponding discrete frequency $\omega_h(k_n)$.

\subsection{Perturbative electromagnetic update}

The first-order electromagnetic fields are advanced by a Yee-like leapfrog update inserted around the native OSIRIS particle and field advance. The perturbative electric field is updated from the curl of the perturbative magnetic field, the first-order current, and the axion source term. In schematic finite-difference form, the electric half-step is
\begin{equation}
  \E_1 \leftarrow \E_1
  + \frac{\dt}{2}\left[\grad_h\times\B_1 - \J_1 + \mathbf{S}_a\right],
  \label{eq:e1-update}
\end{equation}
where the implemented axion source is the discrete analogue of
\begin{equation}
  \mathbf{S}_a = \ag\left[(\partial_t a)\B_0-\E_0\times\grad a\right].
  \label{eq:source-continuum}
\end{equation}
The perturbative magnetic half-step is
\begin{equation}
  \B_1 \leftarrow \B_1 - \frac{\dt}{2}\grad_h\times\E_1.
  \label{eq:b1-update}
\end{equation}
Equations~(\ref{eq:e1-update})--(\ref{eq:b1-update}) are the OSIRIS realization of the first-order Maxwell system in Eqs.~(\ref{eq:fps-faraday}) and (\ref{eq:fps-ampere}). The exact component placement follows the existing OSIRIS finite-difference and mesh-field conventions.

In one spatial dimension, the source routines make the cross-product structure especially transparent. With gradients only along $x$, the code-level source components are
\begin{align}
  S_x &= \ag(\partial_t a)B_x,\label{eq:sx-1d}\\
  S_y &= \ag\left[(\partial_t a)B_y-E_z\partial_x a\right],\label{eq:sy-1d}\\
  S_z &= \ag\left[(\partial_t a)B_z+E_y\partial_x a\right].\label{eq:sz-1d}
\end{align}
The two- and three-dimensional branches implement the same vector expression with the required local averages to place $\partial_t a$, $\grad a$, $\E_0$, and $\B_0$ at the correct field-component locations.

\subsection{Particle field-perturbation splitting}

The OSIRIS implementation also carries the first-order electromagnetic response into the particle push. Following the FPS construction of Ref.~\cite{An2024}, each particle has a zeroth-order position and momentum and a first-order correction,
\begin{equation}
  \x = \x_0+\x_1,
  \qquad
  \p = \p_0+\p_1.
  \label{eq:particle-split}
\end{equation}
Using the mass-normalized momentum $\p=\mathbf{p}_{\rm phys}/(m_sc)$, the FPS particle equations for a particle of species $s$ are
\begin{align}
  \frac{d\x_1}{dt} &= \vvec_1,\label{eq:fps-x1}\\
  \frac{d\p_1}{dt} &= \frac{q_s}{m_s}\left(\E_p+\vvec_0\times\B_p+\vvec_1\times\B_{\rm nop}\right),
  \label{eq:fps-p1}
\end{align}
where $q_s$ and $m_s$ are the species charge and mass, $c$ is the speed of light, $\E_p$ and $\B_p$ are the effective perturbed fields acting on the particle, and $\B_{\rm nop}$ is the unperturbed magnetic field \cite{An2024}.

In OSIRIS, each particle species carries additional first-order
position and momentum information when FPS is enabled. During hybrid
initialization, this first-order state is enabled for all species.
For the FPS field interpolation, a shadow particle position displaced
by $\x_1$ is constructed. The background fields are evaluated at both
the base position $\x_0$ and the shadow position $\x_0+\x_1$, while the
perturbative fields are evaluated at the base position. The effective
first-order fields used in the pusher are therefore
\begin{align}
  \E_p &=
  \E_1(\x_0)
  +\left[
    \E_0(\x_0+\x_1)-\E_0(\x_0)
  \right],
  \label{eq:ep-fps}\\
  \B_p &=
  \B_1(\x_0)
  +\left[
    \B_0(\x_0+\x_1)-\B_0(\x_0)
  \right],
  \label{eq:bp-fps}
\end{align}
with
\begin{equation}
  \B_{\rm nop}=\B_0(\x_0).
  \label{eq:bnop-fps}
\end{equation}
The bracketed terms account for the change in the background field
sampled by the first-order displacement of the particle, while
$\E_1$ and $\B_1$ provide the explicitly evolved axion-induced field
perturbations. The first-order velocity is obtained from the derivative of $\vvec=\p/\gamma$ with respect to momentum. In the normalized variables used by the pusher, the code evaluates
\begin{equation}
  \vvec_1 = \frac{\p_1}{\gamma_0}
  -\p_0\frac{\p_0\cdot\p_1}{\gamma_0^3},
  \label{eq:v1-code}
\end{equation}
where $\gamma_0=(1+|\p_0|^2)^{1/2}$ in the usual OSIRIS relativistic normalization. The first-order momentum update then has the code-level form
\begin{equation}
  \p_1^{q+1}=\p_1^q+\frac{q_s\dt}{m_s}
  \left[\E_p+\vvec_0\times\B_p+\vvec_1\times\B_0\right],
  \label{eq:p1-code-update}
\end{equation}
where the superscript $q$ is the discrete timestep index introduced above. The first-order position is advanced using the updated first-order velocity and the same coordinate normalization conventions as the standard OSIRIS particle advance.

\subsection{Perturbative current deposition}

The perturbative current $\J_1$ is required in Eq.~(\ref{eq:e1-update}). The FPS method expands the current deposition about the zeroth-order particle trajectory \cite{An2024}. In OSIRIS, this is implemented using a particularly direct numerical construction: the code deposits two currents using the native OSIRIS current-deposition routine and subtracts them.

For each particle, the base trajectory uses $(\x_0,\p_0)$ and the shadow trajectory uses $(\x_0+\x_1,\p_0+\p_1)$. The code deposits a base current $\J_{\rm base}$ and a shadow current $\J_{\rm shadow}$ with the same particle shape, grid staggering, and deposition machinery as ordinary OSIRIS. The perturbative current is then accumulated as
\begin{equation}
  \J_1 = \J_{\rm shadow}-\J_{\rm base}.
  \label{eq:j1-shadow-base}
\end{equation}
This avoids introducing an inconsistent hand-coded derivative of the shape function. It also guarantees that the perturbative current is represented on the same mesh and with the same deposition conventions as the native current. This is an important part of the implementation because the perturbative Maxwell update should see a current that is numerically compatible with the current-deposition stencil used by the underlying PIC code.

The present FPS particle-current path is implemented for linear particle interpolation in Cartesian geometry. If a simulation attempts to use the FPS path with a different interpolation order, the code aborts rather than silently running an unsupported configuration.

\subsection{Timestep ordering}

The hybrid timestep is organized so that the scalar field, perturbative electromagnetic fields, particles, and currents are staggered consistently with the inherited OSIRIS leapfrog loop. A single iteration can be summarized as follows. First, OSIRIS performs the usual bookkeeping, laser injection, moving-window operations if enabled, and native field and boundary updates. The axion-field boundaries are then updated. The first axion-electromagnetic half-step applies the backreaction source to $\E_1$ for $\dt/2$, advances $\B_1$ for $\dt/2$, and advances the scalar axion field for a full timestep using the driven Klein-Gordon update.

The code then enters the particle stage. The FPS context is activated, the ordinary particles are pushed, the native current $\J_0$ is deposited, and the perturbative current $\J_1$ is constructed through the shadow-minus-base deposition described above. Both ordinary and perturbative currents are then updated and smoothed according to their respective OSIRIS pathways. The native OSIRIS electromagnetic fields are advanced using the ordinary current. Finally, the second perturbative electromagnetic half-step advances $\B_1$ for another $\dt/2$ and applies the second $\E_1$ half-step using the freshly deposited first-order current.

This ordering matters because the scalar field requires a centered $\E\cdot\B$ source, the perturbative electromagnetic response requires centered values of $a$, $\partial_t a$, and $\J_1$, and the particles require both zeroth-order and first-order fields. The chosen sequence is therefore a practical staggered coupling between the axion solver and the native OSIRIS PIC loop. It preserves the ordinary OSIRIS update structure while inserting the coherent axion-field and perturbative electromagnetic updates at compatible times.

\subsection{Boundary conditions}

The axion boundary module maps user-provided boundary choices onto three internal categories: periodic, reflecting, and open. The periodic fill copies guard-cell values from the opposite side of the domain. The reflecting fill mirrors interior values and corresponds to a zero-normal-gradient scalar boundary for the scalar field. The open fill uses a simple extrapolating open/outflow-style guard-cell fill. This should be distinguished from the full perfectly matched layer (PML) treatment of Ref.~\cite{An2024}, which implements reflecting, periodic, outflow, and PML boundary conditions for the EPOCH axion-field solver. In the present OSIRIS implementation, absorbing-boundary-style input names are interpreted through the simpler open boundary treatment rather than through a separate absorbing-layer evolution equation.

The same boundary module applies the selected fill to scalar fields, scalar diagnostics, and perturbative electromagnetic vector fields. The hybrid loop updates axion boundaries before the first perturbative half-step and again after the second perturbative update. Additional local boundary fills are applied inside the solver around perturbative field updates and source evaluations. For the standing-wave benchmark used here, periodic longitudinal boundaries are the cleanest choice because they maintain well-defined pump and sideband Fourier modes. For finite-pulse propagation studies, open boundaries may be useful, but their residual reflection should be checked directly in the relevant simulation geometry.

\subsection{Diagnostics and normalization}

The axion diagnostics are implemented through the standard OSIRIS field-diagnostic framework. The available report quantities include the centered scalar field $a$, the centered time derivative $\partial_t a$, the two stored scalar half-time levels, the three perturbative electric-field components, the three perturbative magnetic-field components, the centered $\E\cdot\B$ source, and the magnitude of the perturbative electromagnetic source. These quantities are written under the normal OSIRIS field-diagnostic structure and can be processed with the same HDF5-based analysis workflow used for ordinary OSIRIS field outputs.

The diagnostics are designed to expose both the source and the response of the axion-electrodynamics system. The $\E\cdot\B$ diagnostic measures the scalar-field drive. The $a$ and $\partial_t a$ diagnostics allow Fourier-mode amplitudes, mode energies, and growth rates to be reconstructed. The perturbative electromagnetic diagnostics $\E_1$ and $\B_1$ measure the axion-induced electromagnetic response and provide the quantities needed to study pump or probe polarization changes. Additional printed diagnostics accumulate scalar kinetic, gradient, and mass-energy-like contributions, perturbative electromagnetic magnitudes, and source norms.

OSIRIS advances the equations in its native normalized units. The axion diagnostic metadata follows the standard OSIRIS normalization, in which lengths and times are expressed in units of $c/\omega_{\rm ref}$ and $1/\omega_{\rm ref}$, respectively. In the usual plasma formulation $\omega_{\rm ref}$ is identified with the reference plasma frequency $\omega_p$; in the vacuum benchmark considered here it simply sets the arbitrary physical scale of the normalized problem. The deck-level axion mass, axion-photon coupling, laser amplitude, laser frequency, and mode number therefore need to be interpreted in this code unit system. For comparison with the standing-wave instability, the useful dimensionless combinations are $\mu=m_a/\omega_0$, $\alpha=g_{a\gamma\gamma}A_L$, and $K=k\omega_0/m_a^2$, as introduced in Sec.~\ref{sec:two-beam-theory}. For a periodic box, Eq.~(\ref{eq:box-mode}) gives the link between an input mode number and the corresponding simulation wavenumber.

\subsection{Scope of the present implementation}

The current implementation is intended for coherent, weakly
perturbative axion-electrodynamics simulations. The perturbative
electromagnetic response is evolved as a first-order field, which is
appropriate for the early-time growth of the parametric instability studied here
and for weak axion-induced electromagnetic modulations. Strongly
nonlinear regimes in which the perturbative field becomes comparable
to the background field may require an extension beyond the present
first-order splitting.

The current FPS particle path supports linear particle interpolation. The axion boundary implementation supports periodic, reflecting, and extrapolating open fills, but does not currently implement a separate absorbing-layer PML evolution for the scalar field. These limitations define the present operating scope of the implementation. The vacuum benchmark presented below exercises the scalar-field solver, the perturbative electromagnetic update, and their coherent two-way coupling, but does not involve kinetic particles. The FPS particle-push and perturbative-current pathways are therefore not tested by the present benchmark and will require separate validation in kinetic plasma configurations.

\section{Benchmarking results and physical interpretation}
\label{sec:benchmark-results}

\subsection{Simulation configuration and analysis}

The instability is tested in one-dimensional vacuum simulations using the coherent-field OSIRIS module. The periodic domain has length
\begin{equation}
  L=5.0\times10^4,
\end{equation}
and is resolved with $N_x=100000$ cells, giving $\Delta x=0.5$. The timestep is $\Delta t=0.25$. The pump mode number is $N_0=2000$, so that
\begin{equation}
  k_0=\omega_0=\frac{2\pi N_0}{L}=0.2513274123.
\end{equation}
Two equal counter-propagating pumps are launched with the same linear polarization and per-beam vector-potential amplitude $A_L=0.3$, producing the standing-wave background of Eq.~(\ref{eq:pump-potential}). The axion mass is fixed by
\begin{equation}
  \frac{m_a}{\omega_0}=0.06,
  \qquad
  m_a=0.01507964474.
\end{equation}
The coupling is then varied over $g_{a\gamma\gamma}=0.60$, $0.80$, and $1.00$, with all other physical and numerical parameters held fixed. Each production run extends to $t=9000$, and field diagnostics are written every 20 code-time units.

A weak broadband axion seed is imposed over absolute Fourier modes $N_a=1984,\ldots,2016$, corresponding to envelope sectors $\ell=0,\ldots,16$ around the pump mode $N_0=2000$. The initial axion amplitude is $10^{-12}$ per seeded component, with an incommensurate phase increment across the band to avoid coherent spatial addition. No perturbative electromagnetic daughter is seeded; it is generated dynamically by the axion--pump coupling. Because the standing pump couples the pair $N_0-\ell$ and $N_0+\ell$, the two Fourier components are treated as one Floquet sector when extracting a growth rate.

For each physical axion Fourier mode $N$, with $k_N=2\pi N/L$ and $\omega_N=\sqrt{k_N^2+m_a^2}$, we construct a phase-independent modal amplitude from the Fourier coefficients of the scalar field and its time derivative,
\begin{equation}
  \mathcal{A}_N(t)=
  \left[
    \left(2|\widetilde{a}_N(t)|\right)^2
    +\left(
      \frac{2|\widetilde{\partial_t a}_N(t)|}{\omega_N}
    \right)^2
  \right]^{1/2}.
  \label{eq:modal-amplitude}
\end{equation}
Here the factor of two converts the positive-frequency coefficient of the real-valued spatial Fourier transform to the corresponding real-mode amplitude. The symmetry-related modes are then combined into the Floquet-sector amplitude
\begin{equation}
  A_\ell(t)=
  \begin{cases}
    \mathcal{A}_{N_0}(t), & \ell=0,\\[1mm]
    \left[
      \mathcal{A}_{N_0-\ell}^2(t)
      +\mathcal{A}_{N_0+\ell}^2(t)
    \right]^{1/2}, & \ell>0.
  \end{cases}
  \label{eq:sector-amplitude}
\end{equation}
For each $\ell$, $\ln A_\ell(t)$ is fit by linear least squares over the common interval $2500<t<8500$, and the fitted slope defines the simulated growth rate $\Gamma_{\ell,\mathrm{sim}}$. A sector is classified as exponentially unstable when $\Gamma_{\ell,\mathrm{sim}}>0$ and the log-linear fit has coefficient of determination $R^2\ge0.995$. Stable or oscillatory sectors are reported as $\Gamma_{\ell,\mathrm{sim}}=0$ rather than assigning physical significance to a small regression slope across an oscillatory signal. The theoretical curves are obtained from the converged Floquet system in Eqs.~(\ref{eq:floquet-b})--(\ref{eq:floquet-a}); their positive imaginary eigenvalue is denoted $\Gamma_{\ell,\mathrm{th}}$. The dashed cutoff markers in Fig.~\ref{fig:mode-cutoff} show the dominant-branch instability boundary.

For the three couplings, the reduced parameter
$r=(g_{a\gamma\gamma}A_L\omega_0/m_a)^2$ takes the values $9$, $16$, and $25$. Equation~(\ref{eq:gamma0}) therefore predicts zero-envelope growth rates
\begin{equation}
  \Gamma_0=
  8.46\times10^{-4},\quad
  1.20\times10^{-3},\quad
  1.53\times10^{-3},
\end{equation}
for $g_{a\gamma\gamma}=0.60$, $0.80$, and $1.00$, respectively. The corresponding reduced-model cutoffs from Eq.~(\ref{eq:ell-cutoff}) are
\begin{equation}
  \ell_+=8.44,\qquad10.80,\qquad13.23.
  \label{eq:benchmark-cutoffs}
\end{equation}
These predictions require no fitted simulation parameter.

\subsection{Growth-rate benchmark}

\begin{figure*}[!t]
  \centering
  \includegraphics[width=0.82\textwidth]{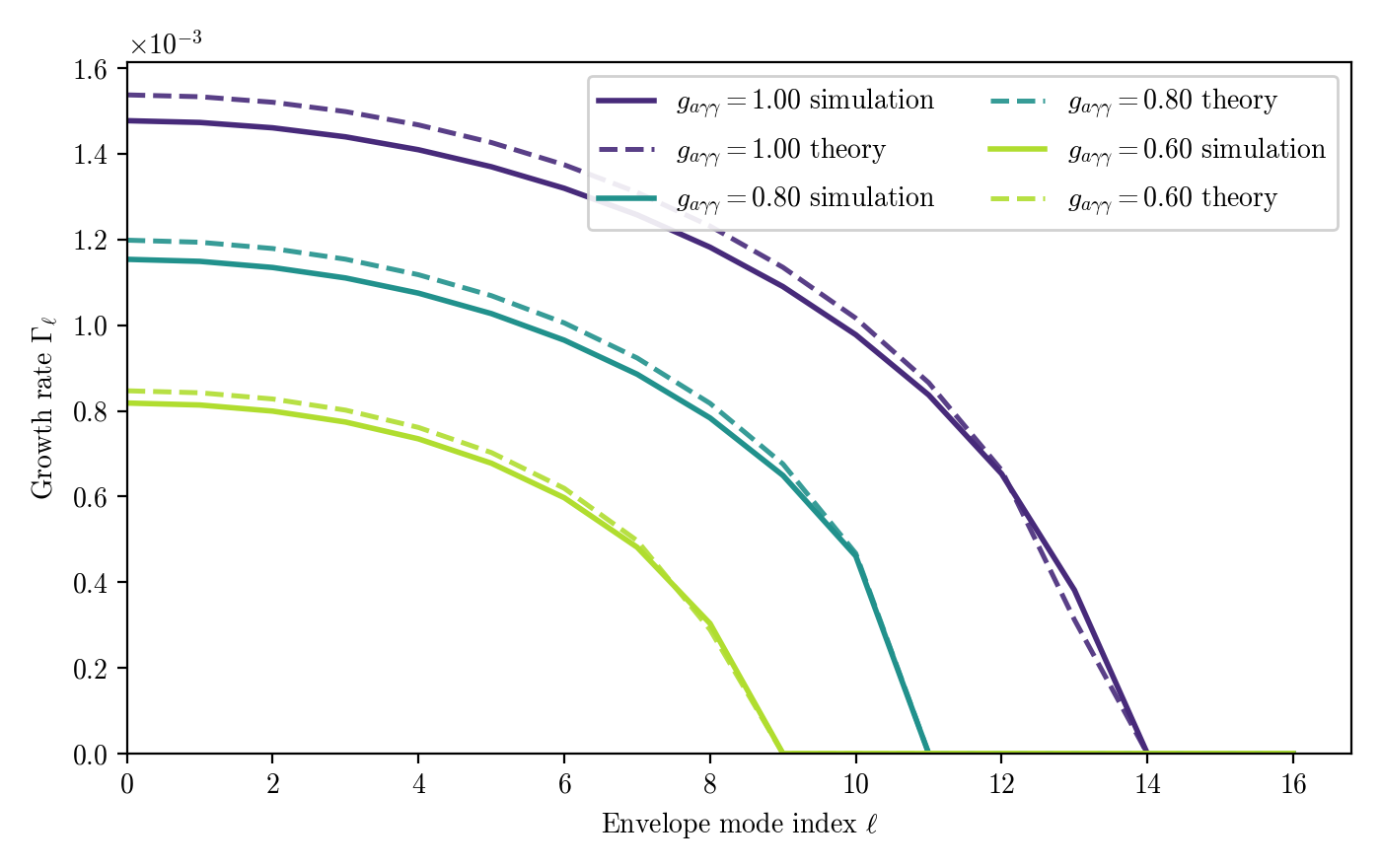}
  \caption{Absolute growth rate as a function of envelope mode index $\ell$ for $g_{a\gamma\gamma}=1.00$, $0.80$, and $0.60$. Solid curves show OSIRIS measurements and dashed curves show the Floquet theory. The simulation reproduces both the coupling dependence of the maximum growth rate and the decrease in $\Gamma_{\ell}$ towards the instability boundary. The plotting scripts were developed with AI assistance; the data points are OSIRIS measurements.}
  \label{fig:growth-benchmark}
\end{figure*}

Figure~\ref{fig:growth-benchmark} compares the mode-resolved OSIRIS growth rates with the Floquet prediction. The strongest growth occurs at the smallest envelope wavenumber and decreases continuously as the edge of the unstable band is approached. Increasing $g_{a\gamma\gamma}$ raises the overall growth rate and broadens the unstable interval, exactly as expected from Eqs.~(\ref{eq:standing-growth}) and (\ref{eq:standing-cutoff}).

For the strongest-coupling case, representative modes well inside the unstable band give $\Gamma_{\rm sim}/\Gamma_{\rm th}\simeq0.96$, where $\Gamma_{\rm sim}$ is the fitted OSIRIS growth rate and $\Gamma_{\rm th}=\operatorname{Im}\omega$ is the corresponding positive Floquet growth rate, over $\ell=0$--$8$, rising towards unity as the cutoff is approached. At $\ell=12$, for example, the simulated and theoretical rates agree to within approximately one per cent. The final point immediately below the cutoff is more sensitive because the theoretical growth rate is small and the finite fitting interval contains fewer e-foldings; this is visible in the $g_{a\gamma\gamma}=1.00$, $\ell=13$ point. The same overall agreement is seen for the $0.80$ and $0.60$ scans.

The rate comparison is a stringent test because the theory curves are obtained from the continuum equations and the pump parameters specified independently in the deck. The simulation is not merely reproducing a qualitative exponential trend: it tracks the full fall of the unstable eigenvalue with envelope mode and its systematic increase with coupling.

\subsection{Coupling-dependent cutoff}

\begin{figure*}[!t]
  \centering
  \includegraphics[width=0.82\textwidth]{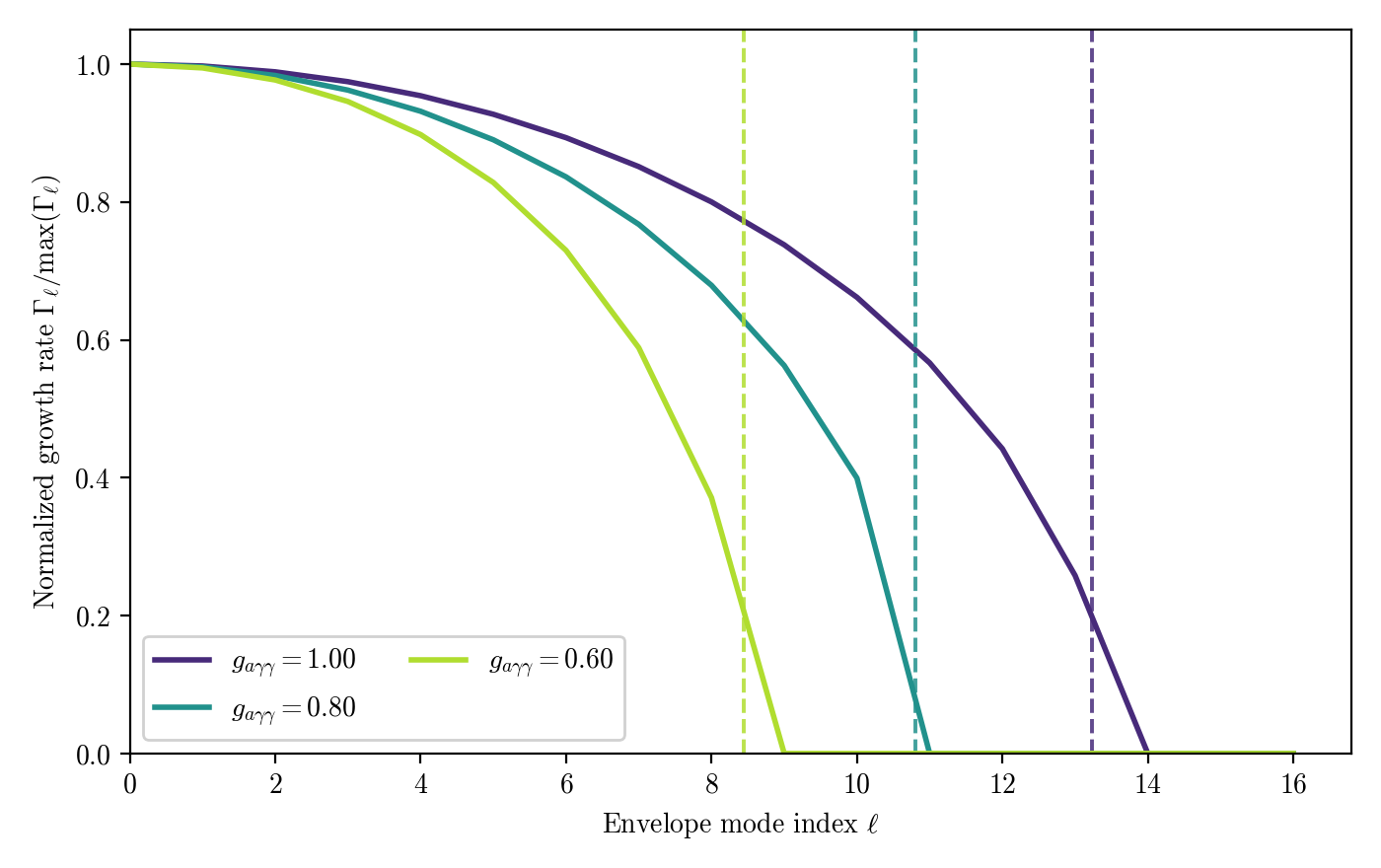}
  \caption{Simulated growth spectra normalized to the maximum growth rate at each coupling. Vertical dashed lines show the theoretical dominant-branch cutoffs. The measured unstable bands terminate at $\ell=8$, $10$, and $13$ for $g_{a\gamma\gamma}=0.60$, $0.80$, and $1.00$, respectively, while the next resolved sectors are stable. The plotting scripts were developed with AI assistance; the data points are OSIRIS measurements.}
  \label{fig:mode-cutoff}
\end{figure*}

The instability boundary is shown more directly in Fig.~\ref{fig:mode-cutoff}. For $g_{a\gamma\gamma}=0.60$, Eq.~(\ref{eq:ell-cutoff}) predicts $\ell_+=8.44$ and the simulations show clean exponential growth through $\ell=8$, with $\ell\ge9$ stable. For $g_{a\gamma\gamma}=0.80$, the predicted boundary is $\ell_+=10.80$ and the observed transition lies between $\ell=10$ and $11$. For $g_{a\gamma\gamma}=1.00$, the prediction is $\ell_+=13.23$ and the transition lies between $\ell=13$ and $14$. Thus all three scans place the cutoff in the unique interval selected by theory at the available integer-mode resolution.

This coupling-dependent movement is especially useful as a numerical benchmark. An unrelated grid or time-integration instability could generate exponential growth, but it would not naturally reproduce the predicted shift of the unstable boundary from $8.44$ to $13.23$ while simultaneously matching the mode-resolved growth spectrum. The result therefore tests the complete feedback loop: the axion source, the perturbative Maxwell response, the Yee-grid centering and interpolation, and the return drive through $\E\cdot\B$.

\subsection{Electromagnetic sidebands of the unstable Floquet mode}

The reduced eigenvector in Eq.~(\ref{eq:unstable-eigenvector}) makes a prediction not contained in the growth rate alone: at zero envelope wavenumber the unstable axion components have spatial harmonics near $\pm k_0$, while the central electromagnetic harmonic $b_{00}$ vanishes. Electromagnetic sidebands at spatial harmonics near $\pm2k_0$ remain allowed and inherit the common Floquet growth exponent. We test this structure using the separately stored axion field and first-order electromagnetic diagnostics in the G060 run ($\ag=0.60$), for which the pump occupies Fourier number $N_0=2000$.

\begin{figure*}[!t]
 \centering
 \includegraphics[width=0.93\textwidth]{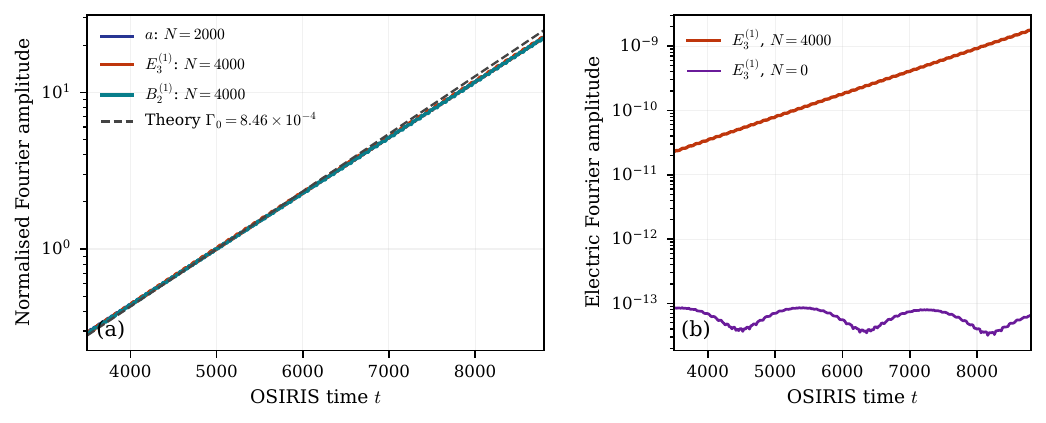}
 \caption{Independent electromagnetic-eigenmode check using the G060 OSIRIS diagnostics. (a) Five-output root-mean-square spatial Fourier amplitudes for the axion at $N=2000$ and the first-order electric and magnetic fields at $N=4000$, each normalized to its value at $t=5000$; the dashed curve is the reduced-model growth $\exp[\Gamma_0(t-5000)]$ with $\Gamma_0=8.46\times10^{-4}$. (b) Absolute spatial Fourier amplitude of the $N=4000$ electric sideband compared with the $N=0$ electric component. The former grows exponentially, while the latter remains small and oscillatory. Fits quoted in the text use $2500<t<8500$. The electric and magnetic quantities are the separately evolved first-order fields, not the background pump. The diagnostic processing and plotting scripts were developed with AI assistance; the field data are from OSIRIS.}
 \label{fig:g060-em-eigenmode}
\end{figure*}

Figure~\ref{fig:g060-em-eigenmode} compares the spatial Fourier amplitudes of the axion field at $N=2000$ and the perturbative electric and magnetic fields at $N=4000=2N_0$. To suppress carrier-scale oscillations, each amplitude is displayed as the root mean square over five neighboring diagnostic times (each output is separated by 20 code-time units). The Fourier-analysis and fitting scripts used for this additional diagnostic test were developed with assistance from OpenAI ChatGPT (GPT-6); the numerical field data themselves are from OSIRIS, and the fitting outputs were cross-checked by reprocessing those diagnostic archives. The three components exhibit nearly identical exponential growth. Fitting their logarithms over $2500<t<8500$, the same interval used for the principal benchmark, gives $\Gamma_a=8.184\times10^{-4}$, $\Gamma_{E_3^{(1)}}=8.241\times10^{-4}$ and $\Gamma_{B_2^{(1)}}=8.210\times10^{-4}$. These results lie close to the reduced zero-envelope prediction $\Gamma_0=8.46\times10^{-4}$. The common growth rate is more significant here than the different raw amplitudes, which correspond to fields with different normalizations and temporal harmonics.

By contrast, the $N=0$ electric component remains small and oscillatory without sustained exponential growth (a log-linear fit has $R^2\simeq0.03$). The vanishing $N=0$ magnetic spatial Fourier coefficient is not an independent test because it follows from the one-dimensional transverse-field representation. The growing $N=4000$ electromagnetic sidebands, together with the non-growing $N=0$ electric mode, provide a diagnostic check of the \emph{mode structure}, distinct from the previously reported axion growth-rate and cutoff benchmarks. The finite-$\ell$ eigenvectors generally permit a nonzero central electromagnetic component.

\subsection{Physical interpretation and scope}

The instability differs qualitatively from the strictly co-linear proposal \cite{Beyer2023}. The equal counter-propagating pumps create a standing electromagnetic background, and the periodic pump coefficients couple a ladder of axion and photon Floquet harmonics. This is not the decay of a single on-shell vacuum photon into a photon and a massive axion, which is kinematically forbidden. Instead, repeated pump--sideband interactions couple distinct axion harmonics via electromagnetic fluctuations. The underlying feedback involves two axion--photon vertices and is second order in the pump--coupling product, even though the unstable eigenfrequency can become approximately linear in that product when the coupling is sufficiently above the $K=0$ threshold. A positive diagonal self-energy correction alone does not determine the stability of this time-periodic coupled system: the off-diagonal Floquet mode couplings must also be retained. The growing solution is thus an axion-dominated parametric sideband instability, with exponentially growing electromagnetic sidebands but no growing central $K=0$ photon harmonic.

The simulations use deliberately large code-level couplings and an idealized periodic vacuum geometry. They are intended as a verification problem rather than as a direct experimental sensitivity estimate. The value of the PIC formulation is that the same corrected equations can subsequently be applied to finite counter-propagating pulses, finite focal spots, imperfect overlap, plasma dispersion, guiding structures, and kinetic plasma backgrounds, where a global Floquet treatment is no longer available. Petawatt laser systems already reach intensities above $10^{23}~\mathrm{W\,cm^{-2}}$ \cite{Yoon2021}, and long plasma-guiding structures have been demonstrated \cite{Picksley2020}; determining whether a realistic configuration accumulates enough gain requires following the finite interaction region directly rather than extrapolating the periodic-box result.

\section{Conclusion}
\label{sec:conclusion}

We have implemented coherent axion electrodynamics in the OSIRIS particle-in-cell framework and used it to benchmark a corrected parametric axion--photon instability. The extension advances a cell-centered Klein--Gordon field and a separately stored first-order electromagnetic response on the native staggered field grid. Field perturbation separation preserves weak axion-induced fields independently of the much larger background laser field, while the FPS particle push and shadow-minus-base current deposition provide a route to kinetic plasma simulations.

The axion--Maxwell equations support a parametric sideband instability when two equal counter-propagating pumps form a standing-wave background. We derived the corresponding Floquet system and a reduced analytic growth rate and cutoff for the dominant envelope branch.

OSIRIS reproduces the theory at three different couplings. For $g_{a\gamma\gamma}=0.60$, $0.80$, and $1.00$, the predicted envelope cutoffs are $\ell_+=8.44$, $10.80$, and $13.23$, while the simulated unstable bands terminate at $\ell=8$, $10$, and $13$, respectively. The absolute growth-rate spectra also closely follow the Floquet roots across the unstable band. Agreement in both quantities provides a stronger benchmark than exponential growth alone. The additional G060 electromagnetic analysis confirms that the growing axion carrier shares its exponent with the $N=4000$ electromagnetic sidebands, whereas the central $N=0$ electric component does not grow; this directly checks the eigenvector structure of the corrected two-way axion--electromagnetic coupling.

The natural next step is to move beyond the periodic standing-wave benchmark to finite counter-propagating pulses and multidimensional geometries, where pulse overlap, diffraction, transverse structure, plasma dispersion, and boundary losses determine the integrated gain. The present result supplies the analytic and numerical reference point required for those studies.

\begin{acknowledgments}
OpenAI ChatGPT (GPT-5.6 and GPT-6) was used as an assistive tool for parts of
the theoretical analysis, symbolic checks of the reduced Floquet determinant,
Python analysis and visualization scripts, and manuscript drafting and editing.
The algebraic results were cross-checked against the original coupled matrix,
and the additional electromagnetic diagnostic fits were cross-checked by
reprocessing OSIRIS data. The authors retain responsibility for the scientific
interpretation and the correctness of the submitted work.
\end{acknowledgments}

\section*{Data availability}
The simulations were performed using the OSIRIS particle-in-cell code,
whose upstream source repository is maintained by the OSIRIS Consortium
\cite{OSIRISGitHub}. The axion-electrodynamics extension developed for this
study is documented in the authors' OSIRIS-Axions repository
\cite{RadfordOsirisAxionsGitHub}. The simulation input decks and
post-processing scripts are intended to be added to the latter repository
after submission. Until then, these materials and the numerical data
underlying the reported figures are available from the corresponding author
upon reasonable request. The complete raw simulation diagnostics have not
been deposited in a public data archive owing to their volume.

\bibliography{references}

\end{document}